\documentclass{article}
\usepackage{fullpage}
\usepackage[american]{babel}
\usepackage{amssymb}
\usepackage{multirow}
\usepackage{graphicx}
\usepackage{pseudocode}

\newcommand{\eqn}[1]{(\ref{#1})}
\newcommand{\beql}[1]{\begin{equation}\label{#1}}
\newcommand{\eeq}{\end{equation}}
\newtheorem{theo}{Theorem}
\newtheorem{defi}{Definiton}

\title{A Simple and Space Efficient Segment Tree Implementation}
\author{Lei Wang and Xiaodong Wang}

\begin{document}
\maketitle

\begin{abstract}
The segment tree is an extremely versatile data structure. In this paper, a new heap based implementation of segment trees is proposed. In such an implementation of segment tree, the structural information associated with the tree nodes can be removed completely. Some primary computational geometry problems such as stabbing counting queries,  measure of union of intervals, and maximum clique size of Intervals are used to demonstrate the efficiency of the new  heap based segment tree implementation. Each interval in a set $S=\{I_1 ,I_2 ,\cdots,I_n\}$  of $n$ intervals can be insert into or delete  from the  heap based segment tree in $O(\log n)$ time. All the primary computational geometry problems can be solved efficiently. 
\end{abstract}

\section{Introduction}
The segment tree structure, originally discovered  by Bentley\cite{1,9,11}, is used as a one-dimensional data structure for intervals whose endpoints are fixed or
known a priori.
The segment tree is very important in solving some primary computational geometry problem because the sets of intervals stored with the nodes can be structured in
any manner convenient for the problem at hand. Therefore, there are many extensions of segment trees that deal with 2-and higher-dimensional objects \cite{2,3,12,13} .
The segment tree can also easily be adapted to stabbing counting queries: report the number of intervals containing the query point. Instead of a list of the intervals is stored in the nodes, an integer representing the number of the intervals is stored. A query with a point is answered by adding the integers on one search path.
Such a segment tree for stabbing counting queries uses only linear storage and queries require $O(\log n)$ time, so it is optimal.
The segment tree structure, can also be useful in finding the measure of a set of intervals. That is, the length of the union of a set of intervals. It can also be used to find the maximum clique of a set of intervals \cite{5,7,8,10}. Segment trees are generally known as semi-dynamic data structures. The new intervals may only be inserted if their endpoints are chosen from a restricted universe. By using a dynamization technique, van Kreveld and Overmars proposed a concatenable version of the segment tree \cite{4,6}. this can be used to answer
the one-dimensional stabbing queries. In addition to the stabbing queries and standard updates (insertion and deletion of segments), the data structure can support split and concatenate operations.

We will discuss the implementation issues on segment tree in this paper. A very simple and space efficient segment tree implementation is presented.

The organization of the paper is as follows.

In the following 4 sections, we describe our presented segment tree implementation.

In Section 2 the preliminary knowledge for presenting our implementation is discussed.
In Section 3 a heap based segment tree implementation is proposed. In such an implementation of segment tree, the structural information associated with the tree nodes can be removed completely.
In Section 4, we discuss a simpler non-recursive implementation of a heap based segment tree.
Some concluding remarks are provided in Section 5.

\section{Preliminaries}
The set $S=\{I_1 ,I_2 ,\cdots,I_n\}$  of $n$ intervals, each of which is represented by $I_i =[l_i ,r_i ],  l_i ,r_i \in R, l_i\leq r_i  $,  is represented by a data array, $D(S)$, whose entries correspond to the end points,  $l_i$ or $r_i $ , and are sorted in non-decreasing order.
This sorted array is denoted $x[0..N], N =2n-1$. That is, $x[0]\leq x[1]\leq \cdots\leq x[N]$.
In the following discussion, the indexes in the range $[0,N]$  are used to refer to the entries in the sorted array $x[0..N]$. A comparison involving a point $q\in R$ and an index $i, 0\leq i\leq N$, is performed in the original domain in $R$. For instance, $q<i$ is interpreted as $q<x[i]$.
Consider the partitioning of the real line induced by $x[0..N]$. The regions of this partitioning are called elementary intervals. Thus, the elementary intervals are, from left to right:
$ (-\infty ,x[0]],(x[0],x[1]],\cdots,(x[N-1],x[N]],(x[N],\infty)$.
That is, the list of elementary intervals consists of half open intervals between two consecutive endpoints $x[i]$ and $x[i+1]$.
The segment tree for the set $x[0..N]$ is a rooted augmented binary search tree, in which each node $v$ is usually associated with some information as shown by \eqn{eq01}.

\beql{eq01}
\left\{\begin{array}{ll}
v.b & \texttt{beginning of interval} ,\\
v.e & \texttt{end of interval} ,\\
v.key & \texttt{split point} ,\\
v.left & \texttt{left pointer} ,\\
v.right & \texttt{right pointer} ,\\
v.aux & \texttt{augmented data structure}.\\
\end{array} \right.
\eeq

Where,  $v.b$ and $v.e$ are used to represent $[v.b,v.e]$, a interval of indexes from $v.b$ to $v.e$.
The key $v.key$ splits the interval $[v.b,v.e]$ into two subintervals, each of which is associated with each child of $v$. The two tree pointers $v.left$ and $v.right$ point
to the left and right subtrees, respectively.  $v.aux $ is an auxiliary pointer, to an augmented data structure.

Given integers $s$ and $t$, with $0\leq s<t\leq N$, the corresponding segment tree $T(s,t)$ can be built recursively as follows.

\begin{pseudocode}[shadowbox]{build}{s,t}
\textbf{Input}:\texttt{interval} [s,t].\\
v\GETS \texttt{newnode()}.\\
v.b\GETS s, v.e\GETS t.\\
v.left\GETS v.right\GETS \texttt{nil}.\\
\IF s+1<t \THEN
\BEGIN
v.key\GETS m\GETS \lfloor (s+t)/2\rfloor.\\
v.left\GETS  \CALL{build}{s,m}.\\
v.right\GETS  \CALL{build}{m,t}.\\
\END\\
\RETURN{v}.
\end{pseudocode}

In the algorithm, a new node $v$ is created first.
The parameters $v.b$ and $v.e$  associated with node $v$ are then set to $s$ and $t$, which define a interval $[v.b,v.e]$, called a standard interval associated with node $v$.
The standard interval associated with a leaf node is also called an elementary interval.

\begin{defi}\label{df1}
Let $b$ and $e$ be two integers and $0\leq b<e\leq N$. A node $v$ in the segment tree $T(0,N)$ is said to be in the canonical covering of the interval $[b,e]$ if its associated standard interval satisfies the property $[v.b,v.e]\subseteq[b,e]$, while that of its parent node does not.
\end{defi}

It is obvious that if a node $v$ is in the canonical covering, then its sibling node $u$,  the node with the same parent node as $v$, is not, for otherwise the common parent node would have
been in the canonical covering. Thus, at each level of the segment tree, there are at most two nodes belong to the canonical covering of a interval $[b,e]$.
Thus, for each interval $[b,e]$, the number of nodes in its canonical covering is at most $\lceil\log(e-b)\rceil+\lfloor\log(e-b)\rfloor-2$.
In other words, a interval $[b,e]$  can be decomposed into at most $\lceil\log(e-b)\rceil+\lfloor\log(e-b)\rfloor-2$ standard intervals.
The segmentation of interval $[b,e]$ is completely specified by the operation that stores
(inserts) $[b,e]$ into the segment tree $T(0,N)$ , that is, by  performing a call to the following algorithm.

\begin{pseudocode}[shadowbox]{insert}{b,e,v}
\textbf{Input}:\texttt{interval} [b,e],\texttt{a node }v\texttt{ of }T(0,N).\\
\IF b\leq v.b \AND e\geq v.e\textbf{ then }
\texttt{assign } [b,e] \texttt{ to } v.
\ELSE
\BEGIN
\IF b<v.key \textbf{ then } \CALL{insert}{b,e,v.left}.\\
\IF v.key<e \textbf{ then } \CALL{insert}{b,e,v.right}.\\
\END
\end{pseudocode}

The insertion of interval $[b,e]$  into segment tree $T(0,N)$ corresponds to a tour in $T(0,N)$, having a general structure. A (possibly empty) initial path, called PIN, from the root to a node $v*$, called the fork, from which two (possibly empty) paths $P_l$ and $P_r$ issue. Either the interval being inserted is allocated entirely to the fork (in which case $P_l$ and $P_r$ are both empty), or all right-children of nodes of $P_l$, which are not on $P_l$, as well as all left-children of nodes of $P_r$, which are not on $P_r$, identify the fragmentation of $[b,e]$.
See Fig.1 for an illustration. In Fig.1, each node has a node number. The node number is assigned to each node as follows. The root node is numbered 1. If a node is numbered $i$, then its left and right child are numbered $2i$ and $2i+1$ respectively. In the insertion of interval $[2,5]$  into segment tree $T(0,13)$,  the initial path from the root to the node 2 is PIN. The node 2 is fork.  The path $P_l$ goes from fork node 2 to node 9, and the path $P_ r$ goes from fork node 2 to node 11. The node 19 is allocated to the interval as a right child of node 9 on the path $P_l$, and the nodes 10 and 22 are  allocated to the interval as a left child of node 5 and 11 respectively on the path $P_r$.

\begin{figure}
\centering
\includegraphics[width=6cm,height=5cm]{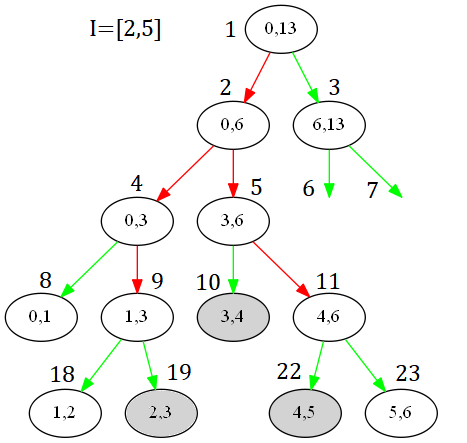}
\caption{Insert $I=[2,5]$ into $T(0,13)$}
\end{figure}

To assign $[b,e]$ to a node $v$  could take different forms, depending upon the requirements of the application.
Frequently all we need to know is the cardinality of the set of intervals allocated to any given node $v$. This can be managed by a single nonegative integer parameter $v.cnt$, denoting this cardinality, so that the allocation of $[b,e]$ to $v$ becomes $v.cnt \leftarrow v.cnt+1$.
In other applications, we need to preserve the identity of the intervals allocated
to a node $v$. Then interval $I =[b,e]$ is inserted into the auxiliary structure associated with node $v$ to indicate that
the standard interval of $v$  is  in the canonical covering of $I$. If the auxiliary structure $v.aux$ associated with node $v$ is an array, the operation assign $[b,e]$ to $v$ can be implemented as $v.aux[i++]=I$.

The insertion algorithm described above can be used to represent a set $S$ of $n$ intervals in a segment tree by
performing the insertion operation $n$ times, one for each interval. As each interval $I$ can have at most $O(\log n)$ nodes in its
canonical covering, and hence we perform at most $O(\log n)$ assign operations for each insertion, the total amount of space
required in the auxiliary data structures reflecting all the nodes in the canonical covering is $O(n\log n)$.

Deletion of an interval $[b,e]$  can be done similarly. The assign operation will be replaced by its corresponding inverse operation remove that removes the interval from the auxiliary structure associated with some canonical covering node.

\begin{pseudocode}[shadowbox]{delete}{b,e,v}
\textbf{Input}:\texttt{interval} [b,e],\texttt{a node }v\texttt{ of }T(0,N).\\
\IF b\leq v.b \AND e\geq v.e\textbf{ then }
\texttt{remove } [b,e] \texttt{ from } v.
\ELSE
\BEGIN
\IF b<v.key \textbf{ then } \CALL{delete}{b,e,v.left}.\\
\IF v.key<e \textbf{ then } \CALL{delete}{b,e,v.right}.\\
\END
\end{pseudocode}

Note that only deletions of previously inserted intervals guarantee correctness.

\section{A Heap Based Implementation}

It is straight forward to see that the segment tree built in the algorithm $\CALL{insert}{0,N,v}$ described above is balanced, and has a height $\lceil \log  N\rceil$.
If a heap is used to store the segment tree nodes, then the structural information associated with the tree nodes can be removed completely.
The heap mentioned above will be defined shortly. It is somewhat different from its definition in the heap sort algorithm where a heap order is defined.
 \begin{defi}\label{df02}
A nearly complete binary tree  or a heap can be defined as follows.
\begin{itemize}
\item The depth of a node $v$ in a binary tree is the length (number of edges) of the path from the root to $v$.
\item The height (or depth) of a binary tree is the maximum depth of any node, or -1 if the tree is empty. Any binary tree can have at most $2^d$ nodes at depth $d$.
\item A complete binary tree of height $h$ is a binary tree which contains exactly $2^d$  nodes at depth $d, 0 \leq d\leq h$.
In this tree, every node at depth less than $h$ has two children. The nodes at depth $h$ are the leaves.
The relationship between $n$ (the number of nodes) and $h$ (the height) is given by $n=1+2+2^2+\cdots+2 ^h=2^{h+1} -1$, and thus $h=\log (n+1)-1$.
\item A nearly complete binary tree of height $h$ is a binary tree of height $h$ in which

(1) There are  $2^d$  nodes at depth  $d$  for $1\leq d\leq h-1$.

(2) The nodes at depth $h$ are as far left as possible.

(3) The relationship between the height and number of nodes in a nearly complete binary tree is given by
$2^h \leq n \leq2^{h+1}-1$, or $h =\lfloor\log n\rfloor$.
\item A heap is a nearly complete binary tree $T$ stored in its breadth-first order as an implicit data structure in an array $A$, where

(1) $A[1]$ is the root of $T$.

(2) The left and right child of $A[i]$ are $A[2i]$ and $A[2i+1]$ respectively.

(3)  The parent of $A[i]$ is $A[\lfloor i/2\rfloor]$.

(4)  If $i$ is odd then$A[i]$ is a right child of its parent $A[\lfloor i/2\rfloor]$, and $A[i-1]$ is its left sibling. If $i$ is even then$A[i]$ is a left child of its parent $A[\lfloor i/2\rfloor]$, and $A[i+1]$ is its right sibling.
\end{itemize}
\end{defi}

\begin{defi}\label{df03}
A heap based segment tree $T(0,N)$ is defined as an array $tree[1..2N-1]$ of tree node elements satisfying the following:
\begin{itemize}
\item The information associated with a tree node $v$ is :
\beql{eq02}
\left\{\begin{array}{ll}
v.cnt & \verb"the number of intervals allocated to node "$v$ ,\\
v.aux & \verb"augmented data structure".\\
\end{array} \right.
\eeq
\item The index $i$ of node $tree[i]$ is called its node number, $1\leq i\leq 2N-1$.
\item The $N$ leaf nodes  corresponding to the $N$ elementary intervals are stored in $tree[N..2N-1]$ in increasing order of their left end point.
In other words, the node $tree[N+i]$ corresponds to the elementary interval $[i,i+1],0\leq i\leq N-1$.
\item The parent node of node $tree[i]$ is $tree[\lfloor i/2\rfloor]$ for all $1<i\leq 2N-1$. The node $tree[1]$ is the root of the heap based segment tree. For each non-leaf node $i, 1\leq i<N$, its left and right children are $2i$, and $2i+1$ respectively.
\end{itemize}
\end{defi}

\begin{figure}
\centering
\includegraphics[width=12cm,height=5cm]{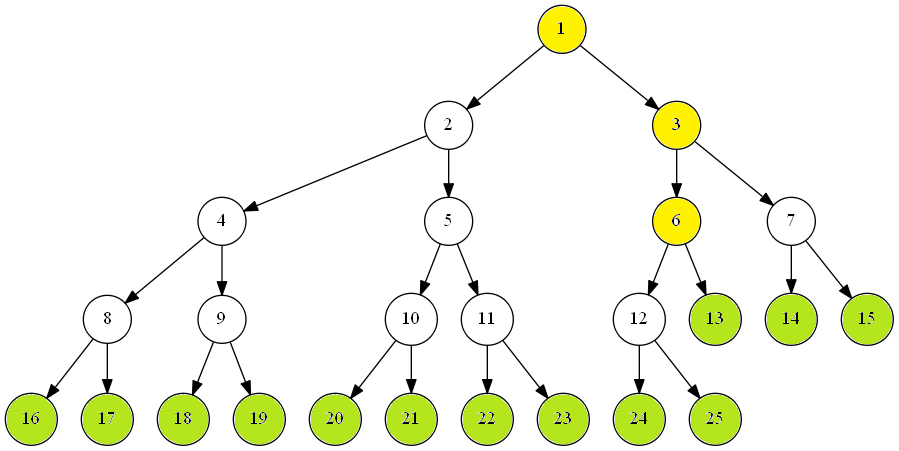}
\caption{A heap based segment tree $T(0,13)$}
\end{figure}

For example, Fig.2 shows the heap based segment tree $T(0,13)$.
It can be seen from Fig.2 and the definition of a heap based segment tree that there are three kinds of nodes in the tree, complete binary tree nodes,  nearly complete  binary tree nodes and leaf nodes.

A complete binary tree node $v$, called a C node, is such a node that the subtree rooted at the node $v$ is a complete binary tree.
A nearly complete binary tree node $v$ (yellow nodes in Fig.2), called a Y node, is such a node that the subtree rooted at the node $v$ is a nearly complete  binary tree node but not a complete binary tree.
A leaf node $v$ (green nodes in Fig.2), corresponds to a leaf of the tree.  The elementary interval $[i,i+1],0\leq i\leq N-1$ is associated with the $N$ leaf nodes numbered $N,N+1,\cdots,2N-1$. The $N$ leaf nodes are also C nodes. There are a total of $2N-1$ nodes in $T(0,N)$, where $N$ leaf nodes and $N-1$ non-leaf nodes. Furthermore, these 3 kinds of nodes satisfy with the following properties.
\begin{theo}\label{th2}
Let $T(0,N)$ be a heap based segment tree, and its nodes are  stored in array $tree[1..2N-1]$ by definition 2, then
\begin{enumerate}
\item[(1)] If node $x$ is a C node, then the high of the subtree rooted at $x$ is $h(x)$, and the leftmost and rightmost nodes of the subtree rooted at $x$ are $l(x)$ and $r(x)$ respectively , and thus the standard interval  associated with node $x$ is $[l(x)-N,r(x)-N+1]$, where
\beql{eq03}
\left\{\begin{array}{ll}
h(x)=\lceil\log(N/x)\rceil &\\
l(x)=x2^{h(x)} & \\
r(x)=(x+1)2^{h(x)}-1& \\
\end{array} \right.
\eeq
\item[(2)]  Let $t(N)$ be the number of trailing zeros of $N$ in its binary expression, then the lowest Y node of  $T(0,N)$ is the node $y(N)=\lfloor N/2^{1+t(N)}\rfloor $. All of the Y nodes of  $T(0,N)$ are  on the path from the root node 1 to node $y(N)$.
\end{enumerate}

\end{theo}

\textbf{Proof}
\begin{enumerate}
\item[(1)] Let the leftmost leaf node of the subtree rooted at node $x$ be $l(x)$. It is readily seen that  $l(x)=2^kx$, where $k=h(x)$ is a nonnegative integer such that $N\leq 2^kx<2N$. It follows that $N/x\leq 2^k<2N/x$, and thus $\log(N/x)\leq k<1+\log(N/x)$.  Therefore, $k=h(x)=\lceil\log(N/x)\rceil$, and $l(x)=x2^{h(x)}$. Since $l(x)$ is a leaf node, its  associated elementary interval is $[l(x)-N,l(x)-N+1]$ by definition 2. Therefore, the left end of the standard interval  associated with node $l(x)$ is $l(x)-N$.
    It is clear that the subtree rooted at node $x$ has $2^{h(x)}$ leaf nodes. It follows that the rightmost leaf node of the subtree rooted at node $x$ must be the node $r(x)=l(x)+2^{h(x)}-1=(x+1)2^{h(x)}-1$. The elementary interval associated with $r(x)$ is then $[(x+1)2^{h(x)}-1-N,(x+1)2^{h(x)}-N]$ by definition 2. It follows that the right end of the standard interval  associated with node $r(x)$ is $(x+1)2^{h(x)}-N=r(x)-N+1$.
\item[(2)]  In the case of $N=2^k$, the segment tree $T(0,N)$ is a complete binary tree of height $k$, and $t(N)=k$.  It follows that $y(N)=0$, and thus there is no Y node in $T(0,N)$. The claim is true for this trivial case.

In the general cases of $N<2^{h(N)}$, the segment tree $T(0,N)$ is a  nearly complete binary tree of height $h(N)$, as shown in Fig. 3.
\begin{figure}
\centering
\includegraphics[width=8cm,height=4cm]{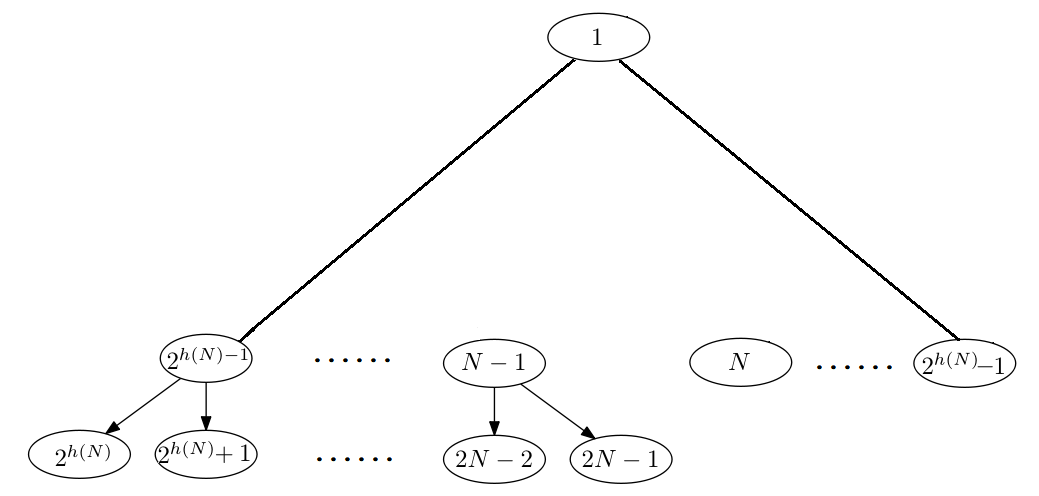}
\caption{A heap based segment tree $T(0,N)$}
\end{figure}
The nodes on the left spine of the tree are numbered $1,2,2^2,\cdots, 2^{h(N)}$, and the nodes on the right spine of the tree are numbered $1,2^2-1,\cdots, 2^{h(N)}-1$. The leaves are distributed at depths $h(N)$ and $h(N)-1$. It is clear that a node $x$ is a Y node if and only if it has a leaf $u$ at  depth $h(N)$ and a leaf $v$ at  depth $h(N)-1$.  The parent of $x$ contains these two leaves either, and thus it is also a Y node. It follows by induction that the nodes on the path from root 1 to $x$ are all Y nodes. Since the node $2N-1$ is the rightmost leaf node at depth  $h(N)$, and the node $N$ is the leftmost leaf node at depth  $h(N)-1$, the Y node $x$ contains leaf nodes $N$ and $2N-1$. Let node $y(N)$ be the lowest common ancestor of  nodes $N$ and $2N-1$. It follows that the Y nodes of the segment tree $T(0,N)$ are all on the path from  root 1 to $y(N)$. It follows from node $N-1$ is the parent node of $2N-1$ that $y(N)$ is also the lowest common ancestor of  nodes $N$ and $N-1$. It is readily seen that if node $u$ is the parent node of $v$, then  $u$ is exactly $v$ shift right 1 bit in its binary expression. It follows that $y(N)$ is exactly the longest common prefix of $N$ and $N-1$ in their binary expression. Let $t(N)$ be the number of trailing zeros of $N$ and $n=\lfloor\log N\rfloor$, then the number of trailing ones of  $N-1$ is also  $t(N)$, and the numbers $N$ and $N-1$ in their binary expression must be
\beql{eq04}
\left\{\begin{array}{ll}
(N)_2=(b_n,b_{n-1},\cdots,,b_{t(N)+1},1,{\overbrace{0,\cdots,0}^{t(N)}}) &\\
(N-1)_2=(b_n,b_{n-1},\cdots,,b_{t(N)+1},0,{\overbrace{1,\cdots,1}^{t(N)}}) &\\
\end{array} \right.
\eeq

It follows that $(y(N))_2=(b_n,b_{n-1},\cdots,b_{t(N)+1})$. In other words, $$y(N)=\lfloor N/2^{1+t(N)}\rfloor =\lfloor (N-1)/2^{1+t(N)}\rfloor.$$
\end{enumerate}

The proof is complete.$\blacksquare$

It follows from Theorem 1 that if node $x$ is a Y node, then its rightmost node is the node $$r'(x)=\lfloor r(x)/2\rfloor=\lfloor ((x+1)2^{h(x)}-1)/2\rfloor=(x+1)2^{h(x)-1}-1$$
In this case,  the standard interval  associated with node $x$ is no longer a single interval, but usually two separated  intervals $[0,r'(x)-N+1]$ and $[l(x)-N,N]$.

For example, int the case of $N=13$ (see Fig. 2), the three C nodes of $T(0,N)$ are 1,3 and 6. It is readily seen that $[l(1)=16,l(3)=l(6)=24$, and $[r'(1)=r'(3)=15,r'(6)=13$.
The interval  associated with nodes 1,3 and 6 are $\{[0,3],[3,13]\}$, $\{[0,3],[11,13]\}$ and $\{[0,1],[11,13]\}$ respectively.

It follows from Theorem 1 that $y(N)$ is a key number for all C nodes. If $x$ is a C node, then it is a prefix of $y(N)$ in binary expression. It follows that $x$ is a C node if and only if
\beql{eq05}
x=\lfloor y(N)/2^{\lfloor\log(y(N))\rfloor-\lfloor\log(x)\rfloor}\rfloor
\eeq
It follows from Theorem 1 that
\beql{eq06}
\left\{\begin{array}{ll}
y(N)=\lfloor N/2^{1+t(N)}\rfloor &\\
t(N)=\log(N\&(-N))&\\
\end{array} \right.
\eeq
Where, $\&$ is a bitwise $\textbf{and}$ operation of two numbers.

Based on Theorem 1, the structural information of any node $x$ in a heap based segment tree $T(0,N)$ can now be computed in $O(1)$ time as follows.

\begin{pseudocode}[shadowbox]{y}{x,N}
\textbf{Input}:\texttt{node } x\texttt{ of }T(0,N).\\
\textbf{Output}:\texttt{the lowest Y node of }T(0,N).\\
t\GETS \log(N\&(-N)).\\
\RETURN{\lfloor N/2^{1+t}\rfloor}.
\end{pseudocode}

\begin{pseudocode}[shadowbox]{c}{x,N}
\textbf{Input}:\texttt{node } x\texttt{ of }T(0,N).\\
\textbf{Output}:\texttt{check for C node of }T(0,N).\\
z\GETS \lfloor \CALL{y}{x,N}/2^{\lfloor\log(\CALL{y}{x,N})\rfloor-\lfloor\log(x)\rfloor}\rfloor.\\
\IF x=z\textbf{ then }\RETURN{\textbf{false}}.
\ELSE \RETURN{\textbf{true}}.
\end{pseudocode}

\begin{pseudocode}[shadowbox]{l}{x,N}
\textbf{Input}:\texttt{node } x\texttt{ of }T(0,N).\\
\textbf{Output}:\texttt{the left end of standard interval}.\\
h\GETS \lceil\log(N/x)\rceil.\\
\RETURN{x2^h-N}.
\end{pseudocode}

\begin{pseudocode}[shadowbox]{r}{x,N}
\textbf{Input}:\texttt{node } x\texttt{ of }T(0,N).\\
\textbf{Output}:\texttt{the right end of standard interval}.\\
h\GETS \lceil\log(N/x)\rceil.\\
\IF \NOT\CALL{c}{x,N}\textbf{ then }h\GETS h-1.\\
\RETURN{(x+1)2^h-N}.
\end{pseudocode}

The following three primary computational geometry problems are used to demonstrate the efficiency of the new  heap based segment tree implementation.
\begin{itemize}
\item \textbf{Stabbing Counting Queries:} Given a set $S=\{I_1 ,I_2 ,\cdots,I_n\}$ of $n$ intervals, each of which is represented by $I_i =[l_i ,r_i ],  l_i ,r_i \in R, l_i\leq r_i  $, and a query point $q$, count all those intervals containing $q$, that is, find a subset $F\subseteq S$ such that $F =\{I_i |l_i \leq q\leq r_i \}$. The problem is to find $|F|$.

\item  \textbf{Measure of Union of Intervals:} Given a set $S=\{I_1 ,I_2 ,\cdots,I_n\}$ of $n$ intervals, the union of $S$ is $U=\bigcup_{I_i\in S}I_i$. The problem is to find  the measure of  $U$.

\item \textbf{Maximum Clique Size of Intervals:} Given a set $S=\{I_1 ,I_2 ,\cdots,I_n \}$ of $n$ intervals, a clique is a subset a subset $C_I\subseteq S$ such that  the common intersection of intervals in $C_I$ is non-empty, and a maximum clique is a clique of maximum size. That is,
$\bigcap_{I_i\in C_I\subseteq S}\neq \emptyset$ and $|C_I|$  is maximized. The problem is to find the maximum size $|C_I|$.
\end{itemize}

The three problems are supposed to be solved simultaneously by using a segment tree to store a set $S=\{I_1 ,I_2 ,\cdots,I_n\}$  of $n$ intervals.
In a heap based segment tree $T(0,N)$, all the structural information are no longer maintained, but only application related information will be associated with the tree nodes.
For the three problems to be solved, the tree nodes are associated with following information.

\beql{eq07}
\left\{\begin{array}{ll}
v.cnt & \texttt{the number of intervals assigned to the node} ,\\
v.uni & \texttt{the measure of the union of intervals assigned to the node} ,\\
v.clq & \texttt{the maximum clique size of intervals assigned to the node}.\\
\end{array} \right.
\eeq

Since there is no structural information to be maintained,  the heap based segment tree is not built explicitly by a procedure like Algorithm 2.1.
The important thing to do is to insert interval set $S$ into the segment tree $T(0,N)$ , that is, by  performing a call to the following algorithm for each interval of $S$.

\begin{pseudocode}[shadowbox]{insert}{b,e,v}
\textbf{Input}:\texttt{interval} [b,e],\texttt{node }v\texttt{ of }T(0,N).\\
\IF \CALL{c}{v,N}
\THEN
\BEGIN
l\GETS \CALL{l}{v,N}.\\
r\GETS \CALL{r}{v,N}.\\
\IF  b>r \OR e\leq l \textbf{ then return}
\ELSEIF b\leq l \AND r\leq e) \textbf{ then } \CALL{change}{v,1}.
\ELSE
 \BEGIN
m\GETS \lfloor(l+r)/2\rfloor.\\
\IF b<m \textbf{ then } \CALL{insert}{b,e,2v}.\\
\IF m<e \textbf{ then } \CALL{insert}{b,e,2v+1}.\\
\END\\
\END
\ELSE
\BEGIN
\CALL{insert}{b,e,2v}.\\
\CALL{insert}{b,e,2v+1}.\\
\END\\
\CALL{update}{v}.
\end{pseudocode}

In above algorithm, a function $\CALL{change}{v,k}$ is used to assign the interval $[b,e]$ to node $v$.

\begin{pseudocode}[shadowbox]{change}{v,k}
\textbf{Input}:\texttt{an integer } k,\texttt{either +1 or -1, and a node  }v\texttt{ of }T(0,N).\\
tree[v].cnt\GETS tree[v].cnt+k.
\end{pseudocode}

The parameter $k$ is 1 in algorithm $\CALL{insert}$, and -1 in algorithm $\CALL{delete}$.

Once the interval is assigned to the node $v$ and its descendent, a function $\CALL{update}{v}$ is invoked to update the information associated with the node $v$ as follows.

\begin{pseudocode}[shadowbox]{update}{v}
\textbf{Input}:\texttt{node }v\texttt{ of }T(0,N).\\
l\GETS \CALL{l}{v,N};
r\GETS \CALL{r}{v,N};
cnt\GETS tree[v].cnt;
ret\GETS 0;\\
\IF cnt>0 \textbf{ then } ret\GETS x[r]-x[l].\\
\COMMENT{a leaf node}\\
\IF r-l=1
\THEN
\BEGIN
 tree[v].clq\GETS cnt.\\
 tree[v].uni\GETS ret.\\
\END\\
\ELSE
 \BEGIN
ul\GETS tree[2v].uni; ur\GETS tree[2v+1].uni;\\
cl\GETS tree[2v].clq; cr\GETS tree[2v+1].clq;\\
tree[v].clq \GETS cnt+\max(cl,cr).\\
\IF cnt>0 \textbf{ then } tree[v].uni \GETS ret
\ELSE tree[v].uni \GETS ul+ur.
\END
\end{pseudocode}

Deletion of an interval $[b,e]$ can be done similarly, except that the parameter $k$ is now replaced by-1 in $\CALL{change}{v,k}$ to remove the interval from some canonical covering node.

\begin{pseudocode}[shadowbox]{delete}{b,e,v}
\textbf{Input}:\texttt{interval} [b,e],\texttt{node  }v\texttt{ of }T(0,N).\\
\IF \CALL{c}{v,N}
\THEN
\BEGIN
l\GETS \CALL{l}{v,N}.\\
r\GETS \CALL{r}{v,N}.\\
\IF  b>r \OR e\leq l \textbf{ then return}
\ELSEIF b\leq l \AND r\leq e) \textbf{ then } \CALL{change}{v,-1}.
\ELSE
 \BEGIN
m\GETS \lfloor(l+r)/2\rfloor.\\
\IF b<m \textbf{ then } \CALL{delete}{b,e,2v}.\\
\IF m<e \textbf{ then } \CALL{delete}{b,e,2v+1}.\\
\END
\END
\ELSE
\BEGIN
\CALL{delete}{b,e,2v}.\\
\CALL{delete}{b,e,2v+1}.\\
\END\\
\CALL{update}{v}.
\end{pseudocode}

It is clear that both of the algorithms $\CALL{change}{v,k}$ and $\CALL{update}{v}$ require $O(1)$ time.
Note that the algorithms $\CALL{insert}{b,e,v}$ and $\CALL{delete}{b,e,v}$ visit at most $O(\log n)$ nodes in the canonical covering of the interval $[b,e]$
and take $O(\log n)$ time. The algorithm $\CALL{insert}{b,e,v}$  must be invoked $n$ times. Therefore, the construction of the heap based segment tree for our purpose requires $O(n\log n)$ time and exactly $2N-1$ units of tree node.

Once all the intervals in $S$ have been inserted into $T(0,N)$, the measure of union of intervals in $S$ is exactly the value stored in ${tree[1].uni}$, and the maximum clique size of intervals in $S$ is exactly the value stored in ${tree[1].clq}$. These values can be found easily in $O(1)$ time as follows.

\begin{pseudocode}[shadowbox]{union}{v}
\textbf{Input}:\texttt{node }v\texttt{ of }T(0,N).\\
\RETURN {tree[v].uni}
\end{pseudocode}

\begin{pseudocode}[shadowbox]{maxclique}{v}
\textbf{Input}:\texttt{node }v\texttt{ of }T(0,N).\\
\RETURN {tree[v].clq}
\end{pseudocode}

To answer a stabbing counting query, a search along a path from root to a leaf is suffice. The search can visit at most $O(\log n)$ nodes, and thus costs $O(\log n)$ time.

\begin{pseudocode}[shadowbox]{stab}{q,v}
\textbf{Input}:\texttt{a query point } q,\texttt{node }v\texttt{ of }T(0,N).\\
c\GETS 0.\\
\IF \CALL{c}{v,N}
\THEN
\BEGIN
l\GETS \CALL{l}{v,N};
r\GETS \CALL{r}{v,N}.\\
\IF q>x[l] \AND q\leq x[r] \textbf{ then } c\GETS  c+tree[v].cnt.\\
\IF r-l>1
\THEN
\BEGIN
m\GETS \lfloor(l+r)/2\rfloor.\\
\IF q<x[m] \textbf{ then }  c\GETS  c+\CALL{stab}{2v}.
\ELSE c\GETS  c+\CALL{stab}{2v+1}.\\
\END
\END
\ELSE
\BEGIN
c\GETS  c+\CALL{stab}{2v}.\\
c\GETS c+\CALL{stab}{2v+1}.\\
\END\\
\RETURN {c}
\end{pseudocode}

\section{A Non-recursive Implementation}

The operations on the heap based segment tree can be realized in a bottom up and non-recursive manner. For example, to insert an interval $[b,e],0\leq b<e\leq N,$ into a heap based segment tree $T(0,N)$, the bottom up searches can be started at two leaf nodes $l=b+N$ and $r=e+N-1$. The two elementary intervals $[b,b+1]$ and  $[e-1,e]$ are associated with the two leaf nodes respectively. In other words,
\beql{eq08}
\left\{\begin{array}{l}
\CALL{l}{l,N}=b\\
\CALL{r}{l,N}=b+1\\
\CALL{l}{r,N}=e-1\\
\CALL{r}{r,N}=e\\
\end{array} \right.
\eeq
In the case of $b+1=e$, the two nodes $l$ and $r$ are coincided, and the only canonical covering node assigned to the interval is found.
In other cases, it is always true that
\beql{eq09}
\left\{\begin{array}{ll}
( \CALL{l}{l,N}, \CALL{r}{l,N} ]\subseteq [ b,e ]&  \\
( \CALL{l}{l,N}, \CALL{r}{l,N} ]\subseteq [ b,e ]&  \\
\end{array} \right.
\eeq

Note that node $l$ is a canonical covering node of $[b,e]$ if and only if it is a right-child of a node on the path $P_l$ (see Fig.1for a reference). Therefore, if $l$ is an odd node, then it is a canonical covering node of $[b,e]$. In this case, node $l+1$ must be an even node. The node $l+1$ cannot be the next canonical covering node of $[b,e]$ unless it is a left-child of a node on the path $P_r$ (see Fig.1for a reference). If the next canonical covering node $v$ of $[b,e]$ is also a right-child of a node on the path $P_l$, then is must be located in the path from the root to node $l+1$ ( see Fig. 4) . The search can now be moved up to the parent node of $l+1$ .
The movement of the node $r$ is totaly symmetric to the movement of the node $r$. The search stops when $l>r$.

\begin{figure}
\centering
\includegraphics[width=5cm,height=4cm]{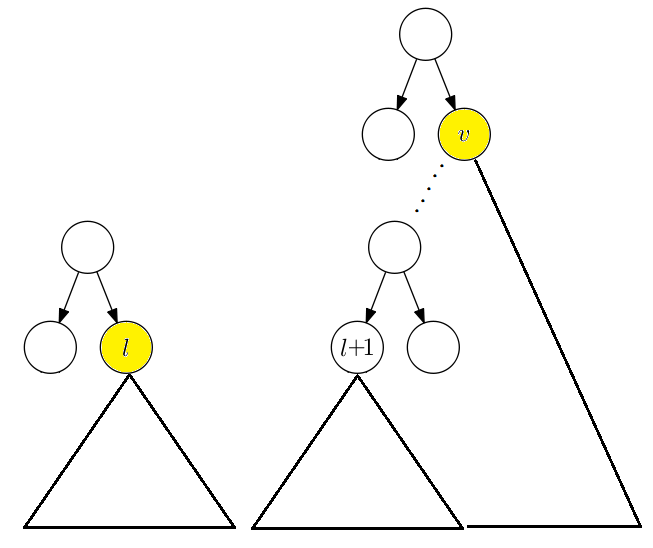}
\caption{search for canonical covering nodes}
\end{figure}

The simple bottom up insertion algorithm can be described as follows.

\begin{pseudocode}[shadowbox]{insert}{b,e}
\textbf{Input}:\texttt{interval} [b,e].\\
l\GETS b+N; r\GETS  e-1+N.\\
s\GETS \lfloor l/2\rfloor; t\GETS \lfloor r/2\rfloor.\\
\WHILE l\leq r \DO
 \BEGIN
\IF \textit{ l}\verb" odd" \THEN
\BEGIN
\CALL{modify}{l,1}.\\
l\GETS (l+1)/2.\\
\END\\
\IF \textit{ r}\verb" even"\THEN
\BEGIN
\CALL{modify}{r,1}.\\
r\GETS \lfloor(r-1)/2\rfloor.\\
\END\\
\END\\
\CALL{pushup}{s}.\\
\CALL{pushup}{t}.
\end{pseudocode}

In above algorithm, $\CALL{modify}{v,k}$ is invoked to assign the interval $[b,e]$ to the canonical covering node $v$.
The parameter $k$ is 1 in algorithm $\CALL{insert}$, and -1 in algorithm $\CALL{delete}$.

\begin{pseudocode}[shadowbox]{modify}{v,k}
\textbf{Input}:\texttt{an integer } k,\texttt{either +1 or -1, and a node  }v\texttt{ of }T(0,N).\\
\CALL{change}{v,k}.\\
\CALL{update}{v}.
\end{pseudocode}

To complete the insertion, the information associated with the nodes on the paths $P_l$ and $P_r$ must be updated also.  The tasks are finished at the end of algorithm by  performing two calls to the following algorithm.

\begin{pseudocode}[shadowbox]{pushup}{v}
\textbf{Input}:\texttt{a node  }v\texttt{ of }T(0,N).\\
\WHILE v>0 \DO
\BEGIN
\CALL{update}{v}.\\
v\GETS \lfloor v/2\rfloor.\\
\END
\end{pseudocode}

Perfectly symmetrical bottom up deletion algorithm can be described as follows.

\begin{pseudocode}[shadowbox]{delete}{b,e}
\textbf{Input}:\texttt{interval} [b,e].\\
l\GETS b+N; r\GETS  e-1+N.\\
s\GETS \lfloor l/2\rfloor; t\GETS \lfloor r/2\rfloor.\\
\WHILE l\leq r \DO
 \BEGIN
\IF \textit{ l}\verb" odd" \THEN
\BEGIN
\CALL{modify}{l,-1}.\\
l\GETS (l+1)/2.\\
\END\\
\IF \textit{ r}\verb" even"\THEN
\BEGIN
\CALL{modify}{r,-1}.\\
r\GETS \lfloor(r-1)/2\rfloor.\\
\END\\
\END\\
\CALL{pushup}{s}.\\
\CALL{pushup}{t}.
\end{pseudocode}

Note that the bottom up and non-recursive algorithms $\CALL{insert}{b,e}$ and $\CALL{delete}{b,e}$ visit at most $O(\log n)$ nodes in the canonical covering of the interval $[b,e]$
and take $O(\log n)$ time. The algorithm $\CALL{pushup}{v}$ requires clearly $O(\log n)$ time. The algorithm $\CALL{insert}{b,e}$  must be invoked $n$ times. Therefore, the non-recursive construction of the heap based segment tree for our purpose requires $O(n\log n)$ time and exactly $2N-1$ units of tree node.

Once all the intervals in $S$ have been inserted into $T(0,N)$, the measure of union of intervals in $S$, and the maximum clique size of intervals in $S$ can be obtained trivially in $O(1)$ time. The stabbing counting query can also be answered top down in $O(\log n)$ time. Alternatively, the stabbing counting query can also be answered in a bottom up manner in $O(\log n)$ time. A binary search is performed firstly in $O(\log n)$ time to find the index $i$ of input array $x$ such that the query  point $q$ is contained in the elementary intervals $(x[i],x[i+1]]$, which is associated with the leaf node $v$. Then, the path from the leaf node $v$ to the root is visited bottom up to count the stabbing number as follows.

\begin{pseudocode}[shadowbox]{stab}{q}
\textbf{Input}:\texttt{a query point } q.\\
a\GETS 0.\\
v\GETS \CALL{bsearch}{q}+N.\\
\WHILE \CALL{c}{v,N} \DO
\BEGIN
a\GETS a+tree[v].cnt;\\
v\GETS \lfloor v/2\rfloor.\\
\END\\
\RETURN {a}
\end{pseudocode}

It is clear that the bottom up stabbing counting algorithm costs $O(\log n)$ time.

\section{Concluding Remarks}
We have suggested a new heap based implementation of segment trees. In such an implementation of segment tree, the structural information associated with the tree nodes can be removed completely. Some primary computational geometry problems such as stabbing counting queries,  measure of union of intervals, and maximum clique size of Intervals are used to demonstrate the efficiency of the new  heap based segment tree implementation. Each interval in a set $S=\{I_1 ,I_2 ,\cdots,I_n\}$  of $n$ intervals can be insert into or delete  from the  heap based segment tree in $O(\log n)$ time. All the primary computational geometry problems can be solved efficiently. Although the  heap based segment tree is also a semi-dynamic data structure. We believe it may hopefully be improved to support split and concatenate operations, since its structure is so simple.

\end{document}